\documentclass[aps,pre,showpacs,showkeys,reprint,floatfix,superscriptaddress]{revtex4-1}

\usepackage{amssymb}
\usepackage{amsmath}
\usepackage{graphicx}
\usepackage{amsbsy}
\usepackage{color}
\usepackage{bm}
\newcommand{\bq}{\begin{eqnarray}}
\newcommand{\eq}{\end{eqnarray}}
\newcommand{\bqn}{\begin{eqnarray*}}
\newcommand{\eqn}{\end{eqnarray*}}
\newcommand{\rr}{\mathbf{r}}

\begin{document}
%%%%%%%%%%%%%%%%%%%%%%%%%%%%%%%%%%%%%%%%%%%%%%%%%%%%%%%%%%%%%%%%%%%%%%%%%%%%%%
%%%%%%%%%%%%%%%%%%%%%%%%%%%%%%%%%%%%%%%%%%%%%%%%%%%%%%%%%%%%%%%%%%%%%%%%%%%%%%
%%%%%%%%%%%%%%%%%%%%%%%%%%%%%%%%%%%%%%%%%%%%%%%%%%%%%%%%%%%%%%%%%%%%%%%%%%%%%%
\title{Two phase coexistence for the hydrogen-helium mixture}

\author{Riccardo Fantoni}
\email{rfantoni@ts.infn.it}
\affiliation{Dipartimento di Scienze Molecolari e Nanosistemi,
  Universit\`a Ca' Foscari Venezia, Calle Larga S. Marta DD2137,
  I-30123 Venezia, Italy} 

%\author{Saverio Moroni}
%\email{moroni@sissa.it}
%\affiliation{SISSA Scuola Internazionale Superiore di Studi Avanzati
%  and DEMOCRITOS National Simulation Center, Istituto Officina dei
%  Materiali del CNR, Via Bonomea 265, I-34136 Trieste, Italy} 

\date{\today}


\begin{abstract}
We use our newly constructed quantum Gibbs ensemble Monte Carlo
algorithm to perform computer experiments for the two phase
coexistence of a hydrogen-helium mixture. Our results are in
quantitative agreement with the experimental results of
C. M. Sneed, W. B. Streett, R. E. Sonntag, and G. J. Van Wylen. 
The difference between our results and the experimental ones is in all
cases less than 15\% relative to the experiment, reducing to less than
5\% in the low helium concentration phase. At the gravitational
inversion between the vapor and the liquid phase, at low temperatures
and high pressures, the quantum effects become relevant. At extremely
low temperature and pressure the first component to show superfluidity
is the helium in the vapor phase. 
\end{abstract}

\maketitle
%%%%%%%%%%%%%%%%%%%%%%%%%%%%%%%%%%%%%%%%%%%%%%%%%%%%%%%%%%%%%%%%%%%%%%%%%%%%%%
\section{Introduction}
%%%%%%%%%%%%%%%%%%%%%%%%%%%%%%%%%%%%%%%%%%%%%%%%%%%%%%%%%%%%%%%%%%%%%%%%%%%%%%
\label{sec:introduction}

Hydrogen and helium are the most abundant elements in the
Universe. They are also the most simple. At ambient
conditions helium is an inert gas with a large band gap. Because of
its low mass and weak inter-atomic interactions, it has fascinating
properties at low temperatures such as superfluidity. The molecular
hydrogen and helium mixture is therefore of special theoretical
importance since it is made by the two lightest elements in nature
which have the lowest critical temperatures. This mixture is found to
make the atmosphere of giant planets like the Jovian and is essential
in stars.  

An important problem to study is the phase coexistence of the fluid
mixture and the determination of its coexistence properties. Some
early experimental studies \cite{Streett1964,Sonntag1964,Sneed1968}
have shown that at coexistence, at low temperature, the mixture shows
a strong asymmetry in species concentrations in the liquid relative to
the vapor phase, with an abundance of helium atoms in the vapor. This 
phenomenon results in the liquid floating above its vapor
\cite{Sneed1968} since helium has approximately twice the molecular
weight of hydrogen. Such experimental coexistence studies has later
been extended at higher temperature and pressure
\cite{Streett1973,Bergh1987} allowing to determine a quite complete
picture for the coexistence phase diagram of this mixture in the
temperature range from $15.5$~K to $360$~K and in the pressure range
from $5$~bars to $75$~kbars. Another interesting issue is whether this
system exhibits fluid-fluid solubility at extremely high pressure 
\cite{Militzer2005,Morales2009,Lorenzen2009,Lorenzen2011,McMahon2012,Morales2013,Soubiran2013},
a situation hard to achieve in the laboratory.

In this work we perform a numerical experiment for the two phase
coexistence problem of the hydrogen-helium mixture at low temperatures
and pressures using the Quantum Gibbs Ensemble Monte Carlo
(QGEMC) method recently devised \cite{Fantoni2014a,Fantoni2014b} to
solve the coexistence of a generic quantum boson fluid where the
particles interact with a given effective pair-potential. We will be
concerned with situations where the absolute temperature, $T$, and the
number density, $\rho_\alpha$, of each one of the two components
$\alpha=a,b$ of mass $m_\alpha$, are such that at least one of the two
components is close to its degeneracy temperature
$(T_D)_\alpha=\rho_\alpha^{2/3}\hbar^2/m_\alpha k_B$, with $k_B$
Boltzmann constant. For temperatures much higher than $\max\{(T_D)_\alpha\}$  
quantum statistic is not very important. This path integral
Monte Carlo simulation enables us to study the quantum fluid mixture
from first principles, leaving the effective pair-potentials between
the two species, the hydrogen molecules and the helium atoms, as the
only source of external information. There are studies on reproducing such
coexistence from an equation of state approach \cite{Wei1996}. Our
QGEMC method is expected to break down at high densities near the
solid phase. Moreover, clearly our approach becomes not anymore
feasible at extremely high pressures when the hydrogen is ionized and
one is left with delocalized metallic electrons
\cite{Militzer2005,Morales2009,Lorenzen2009,Lorenzen2011,McMahon2012,Morales2013,Soubiran2013}.
 
Our binary mixture of particles, of two species labeled by a Greek index,
with coordinates $R\equiv\{\rr_{i_\alpha}|i_\alpha=1,2,\ldots,N_\alpha
~\text{and}~\alpha=a,b\}$, and interacting with a central effective
pair-potential $\phi_{\alpha\beta}(r)$, has a Hamiltonian
\bq \nonumber
\hat{H}&=&-\sum_{\alpha=1}^2\sum_{i_\alpha=1}^{N_\alpha}\lambda_\alpha 
\bm{\nabla}_{i_\alpha}^2+ \\
&&\frac{1}{2}\sum_{\alpha,\beta=1}^2\sum_{i_\alpha,j_\beta}^\prime
\phi_{\alpha\beta}(|\rr_{i_\alpha}-\rr_{j_\beta}|),
\eq
where the prime on the sum symbol indicates that we must exclude the
terms with $i_\alpha=j_\beta$ when $\alpha=\beta$ and
$\lambda_\alpha=\hbar^2/2m_\alpha$.

The density matrix for the binary mixture at equilibrium at an
absolute temperature $T$ is then $\hat{\rho}=e^{-\beta \hat{H}}$ with
$\beta=1/k_BT$. Its coordinate representation $\rho(R,R',\beta)$ can
be expressed as a path ($R(\tau)$) integral in imaginary time ($\tau$)
extending from $R=R(0)$ to $R'=R(\beta)$ \cite{Ceperley1995}. The
many-particle path is made of $N=N_a+N_b$ single-particle world-lines
which constitute the configuration space one needs to sample. Since
the Hamiltonian is symmetric under exchange of like particles we can 
project over the bosonic states by taking
$\rho_B(R,R',\beta)=\sum_{{\cal P}}\rho(R,{\cal P}
R',\beta)/(N_a!N_b!)$ where ${\cal P}$ indicates a permutation of
particles of the same species.  

If we call $\rho$ the number density of the mixture, $x_\alpha$ the molar
concentration of species $\alpha$ ($x_b=1-x_a$), $P=P(T,\rho,x_a)$ the
mixture pressure, and $\mu_\alpha=\mu_\alpha(T,P,x_a)$ the chemical
potential of species $\alpha$, we want to solve the two phase, $I$ and
$II$, coexistence problem 
\bq
\mu_a(T,P,x_a^{(I)})&=&\mu_a(T,P,x_a^{(II)})\\
\mu_b(T,P,x_a^{(I)})&=&\mu_b(T,P,x_a^{(II)})
\eq
for the concentrations, $x_a^{(I)}$ and $x_a^{(II)}$, (and the
densities, $\rho^{(I)}$ and $\rho^{(II)}$) in the two phases. Since
our mixture is not symmetric under exchange of the two species, $a$
and $b$, we expect in general $x_a^{(II)}\neq 1-x_a^{(I)}$.

Our QGEMC algorithm \cite{Fantoni2014b} uses two boxes maintained in
thermal equilibrium at a temperature $T$ and containing the two
different phases. It employs a menu of seven different Monte Carlo
moves: the {\sl volume} move ($q=1$) allows changes in the volumes of
the two boxes assuring the equality of the pressures between the two
phases, the {\sl open-insert} ($q=2$), {\sl close-remove} ($q=3$), and
{\sl advance-recede} ($q=4$) allow the swap of a single-particle
world-line between the two boxes assuring the equality of the chemical
potentials between the two phases, the {\sl swap} ($q=5$) allows to
sample the particles permutations, and the {\sl wiggle} ($q=6$) and
{\sl displace} ($q=7$) to sample the configuration space. 
We thus have a menu of seven, $q=1,2,\ldots,7$, different Monte Carlo
moves with a single random attempt of any one of them occurring with
probability $G_q=g_q/\sum_{q=1}^7 g_q$. 

The paper is organized as follows: In Section \ref{sec:model} we
describe the particular binary mixture studied; in Section
\ref{sec:method} we describe the simulation method employed; in
Section \ref{sec:results} we present our numerical results; Section
\ref{sec:conclusions} is for final remarks.

%%%%%%%%%%%%%%%%%%%%%%%%%%%%%%%%%%%%%%%%%%%%%%%%%%%%%%%%%%%%%%%%%%%%%%%%%%%%%%
\section{The H$_2$-He mixture}
%%%%%%%%%%%%%%%%%%%%%%%%%%%%%%%%%%%%%%%%%%%%%%%%%%%%%%%%%%%%%%%%%%%%%%%%%%%%%%
\label{sec:model}

We consider a binary fluid mixture of molecular hydrogen (H$_2$) and
the isotope helium four ($^4$He), two bosons. We take $1$~\r{A} as
unit of lengths and $k_B$~K 
as unit of energies. We indicate with an asterisk over a quantity its 
reduced adimensional value. We have for the parameter
$\lambda_\alpha=\hbar^2/2m_\alpha$ of the two species
$\alpha=\text{H}_2,\text{$^4$He}$ 
\bq
\lambda^*_{\text{H}_2}&=&12.032,\\
\lambda^*_{\text{He}}&=&6.0596.
\eq
The pair-potential between two helium atoms is the Aziz {\sl et al. }
\cite{Aziz1979} HFDHE2, the one between two hydrogen molecules is the
Silvera {\sl et al.} \cite{Silvera1978}, and the one between a
hydrogen molecule and a helium atom is the Roberts 
\cite{Mason1954,Roberts1963}. All can be put in the following central
form   
\bq \label{pp}
\phi(r)&=&\varepsilon\Phi(x)\\ \nonumber
\Phi(x)&=&\exp(\alpha -\beta x-\gamma x^2)-\\
&&\left(\frac{C_6}{x^6}+\frac{C_8}{x^8}+\frac{C_{10}}{x^{10}}\right)F(x),\\
F(x)&=&\left\{\begin{array}{ll}
\exp[-(D/x-1)^2] & x<D\\
1                & x\geq D
\end{array}\right.,
\eq
where $x=r/r_m$, with $r_m$ the position of the minimum, and the
various parameters are given in Table \ref{tab:pp}. We have
$\phi_{\text{He}\text{He}}^*(r_m)=-10.8$,
$\phi_{\text{H}_2\text{H}_2}^*(r_m)=-34.3$, and
$\phi_{\text{H}_2\text{He}}^*(r_m)=-14.8$. Moreover we have a slight
positive non-additivity:
$[r_m^*]_{\text{H}_2\text{He}}=3.375>([r_m^*]_{\text{He}\text{He}} +
[r_m^*]_{\text{H}_2\text{H}_2})/2=3.189$.  

\begin{figure}[htbp]
\begin{center}
\includegraphics[width=8cm]{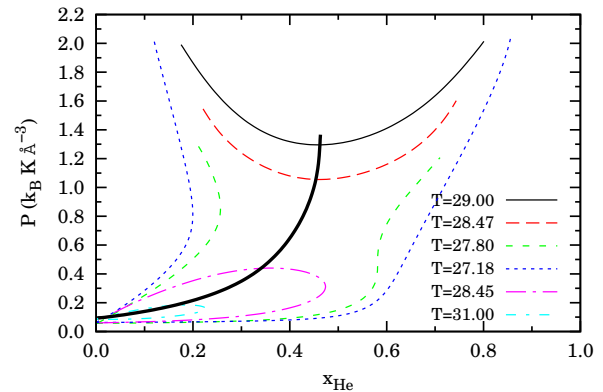}
\end{center}  
\caption{(color online) Schematic pressure-composition phase diagram
  for six isotherms of the hydrogen-helium mixture at low temperatures
  and low pressures as drawn in the experimental work by Streett {\sl
    at al.} \cite{Streett1973}. The thick continuous black line is the
  mixture critical line.}   
\label{fig:pd-streett}
\end{figure}

\begin{figure}[htbp]
\begin{center}
\includegraphics[width=8cm]{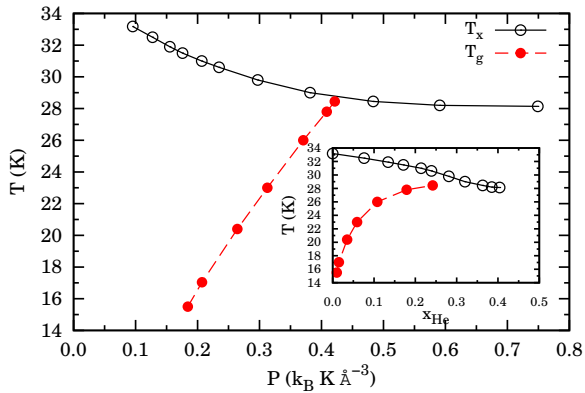}
\end{center}  
\caption{(color online) Reproduction of Fig. 3 of Sneed {\sl et al.}
  \cite{Sneed1968} for the $x-$line and the $g-$line
  (see Eq. (\ref{ginv1})). The inset shows the two lines in the
  temperature-composition plane.}
\label{fig:pd-sneed}
\end{figure}

The experimental coexistence data
\cite{Streett1964,Sneed1968,Streett1973} is given in Table I of the
supplemental material \cite{Supplemental} and represented
schematically in Fig. \ref{fig:pd-streett}.  
For example, the mixture at $T=31~\text{K}$ has a lower 
critical state at $P=0.207~k_B\text{K}\text{\r{A}}^{-3},
x_\text{He}=0.214$ and an upper critical state at
$P=1.96~k_B\text{K}\text{\r{A}}^{-3}, x_\text{He}=0.49$. The set of
all critical states constitutes the $x-$line, $T=T_x(P)$, such
that for $T>T_x$ then $x_\text{He}^{(I)}=x_\text{He}^{(II)}$. The
experimental $x-$line of Sneed {\sl et al.} \cite{Sneed1968}
is shown in Fig. \ref{fig:pd-sneed} for the low temperature and low
pressure mixture. In the figure we also show the experimental line for
the gravitational inversion described in Section \ref{sec:inversion}.  

For temperatures higher than 
the hydrogen critical point $T_{\text{H}_2}=33.19~\text{K}$
($P_{\text{H}_2}=0.094~k_B\text{K}\text{\r{A}}^{-3}$) there is only an
upper critical point \cite{Streett1973}. On the temperature at which
$T_x(P)$ reaches its minimum there is no unanimous consensus among the
various experimental works.

\begin{table*}[htb]
\caption{Pair-potentials parameters: $\phi_\text{pair}$}
\begin{ruledtabular}
\begin{tabular}{llllllllll}
pair & $\varepsilon^*$ & $r_m^*$ & $\alpha$ & $\beta$ & $\gamma$ &
$C_6$ & $C_8$ & $C_{10}$ & $D$\\  
\hline
He-He       & 10.8 & 2.9673 & 13.208 & 13.353 & 0 & 1.3732 &
0.42538 & 0.17810 & 1.2413 \\
H$_2$-H$_2$ & 315778 & 3.41 & 1.713 & 10.098 & 0.41234 & 1.6955$\times 10^{-4}$
& 7.2379$\times 10^{-5}$ & 3.8984$\times 10^{-5}$ & 1.28 \\
H$_2$-He    & 14.76 & 3.375 & 13.035 & 13.22 & 0 & 1.8310 & 0 & 0 & 0.79802 \\
\end{tabular}
\end{ruledtabular}
\label{tab:pp}
\end{table*}

%%%%%%%%%%%%%%%%%%%%%%%%%%%%%%%%%%%%%%%%%%%%%%%%%%%%%%%%%%%%%%%%%%%%%%%%%%%%%%
\section{Simulation method}
%%%%%%%%%%%%%%%%%%%%%%%%%%%%%%%%%%%%%%%%%%%%%%%%%%%%%%%%%%%%%%%%%%%%%%%%%%%%%%
\label{sec:method}

We use our QGEMC method, described in Ref. \cite{Fantoni2014b}, where
we monitor the number densities of the two coexisting phases,
$\rho^{(i)}=N^{(i)}/V^{(i)}=(N^{(i)}_\text{He}+N^{(i)}_{\text{H}_2})/V^{(i)}$
with $i=I,II$, the concentrations of He in the two phases,
$x_\text{He}^{(i)}=N^{(i)}_\text{He}/(N^{(i)}_\text{He}+N^{(i)}_{\text{H}_2})<1$,
and the pressure $P$. We shall
conventionally order $\rho^{(I)}<\rho^{(II)}$ so that $I$ will be the
vapor phase and $II$ the liquid phase, unless $\rho^{(I)}=\rho^{(II)}$,
in which case we have a fluid-fluid phase coexistence. In the
simulation we fix: 
$N=N^{(I)}_\text{He}+N^{(II)}_\text{He}+N^{(I)}_{\text{H}_2}+N^{(II)}_{\text{H}_2}$
with
$N^{(I)}_{\text{H}_2}+N^{(II)}_{\text{H}_2}=\chi[N^{(I)}_\text{He}+N^{(II)}_\text{He}]$
and $V=V^{(I)}+V^{(II)}$. Otherwise   
$N^{(I)}_\text{He},N^{(II)}_\text{He},N^{(I)}_{\text{H}_2},N^{(II)}_{\text{H}_2}$
and $V^{(I)},V^{(II)}$ are allowed to fluctuate keeping
$V^{(I)}+V^{(II)}$ and $N^{(I)}_\text{He}+N^{(II)}_\text{He},
N^{(I)}_{\text{H}_2}+N^{(II)}_{\text{H}_2}$ constants. The Gibbs phase
rule for a two phase 
coexistence of a binary mixture assures that one has two independent
thermodynamic quantities \cite{Landau}. So our control parameters will
be the absolute temperature $T$ and the global number density
$\rho=N/V$ (instead of the pressure as in the experimental
case). As usual a finite $N$ sets the size error for our
calculation. Whereas $\chi>0$ will regulate the size asymmetry
numerical effect so that for
\bq
N^{(I)}_\text{He}&=&\frac{Nx^{(I)}_\text{He}[1-x^{(II)}_\text{He}(1+\chi)]}
{(1+\chi)(x^{(I)}_\text{He}-x^{(II)}_\text{He})}>0,\\
N^{(II)}_\text{He}&=&\frac{N}{1+\chi}-N^{(I)}_\text{He}>0,
\eq
if $x_\text{He}^{(II)}<x_\text{He}^{(I)}$, we must have 
$0<x_\text{He}^{(II)}<1/(1+\chi)<x_\text{He}^{(I)}<1$ and if
$x_\text{He}^{(I)}<x_\text{He}^{(II)}$, then
$0<x_\text{He}^{(I)}<1/(1+\chi)<x_\text{He}^{(II)}<1$. Moreover  
we must also always have $\rho^{(I)}<\rho<\rho^{(II)}$. The initial
condition we chose for our simulations was always as follows:
$\rho^{(I)}=\rho^{(II)}=\rho$ and
$x_\text{He}^{(I)}=x_\text{He}^{(II)}=1/(1+\chi)$. 

Due to the short-range nature of the effective pair-potentials of
Eq. (\ref{pp}) we will approximate, during the simulation, $\phi(r)=0$
for $r>r_\text{cut}\gg [r_m]_{\text{H}_2\text{H}_2}$ (this corresponds
to the truncated {\sl and} not shifted choice in
Ref. \cite{Smit1992}). Where in order to comply with the minimum image 
convention for the potential energy calculation, we make sure
that the conditions $[V^{(i)}]^{1/3}>2r_\text{cut}$, for $i=I,II$, are
always satisfied during the simulation. This approximation is the only
other source of error apart from the size one. The two are related
because for instance in the fluid-fluid coexistence, when
$V^{(I)}\approx V^{(II)}\approx V/2$ during the simulation, we require
$r_\text{cut}\approx(N/2\rho)^{1/3}/2\gg [r_m]_{\text{H}_2\text{H}_2}$
for some given $\rho$.  

The path integral discretization imaginary time step
$\delta\tau=\beta/K$, with $K$ the number of time slices, is chosen so
that $\delta\tau^*=0.002$, which is considered sufficiently small 
to justify the use of the primitive approximation of the inter-action
\cite{Ceperley1995}. The parameters $\bar{M}$, defined in
\cite{Fantoni2014b}, will be called $\bar{M}_q$ for each relevant move
$q$ and the
parameter $\Delta_\Omega$, also defined in \cite{Fantoni2014b}, is
always chosen equal to $0.01$. In order to fulfill detailed balance
we must choose $\bar{M}_2=\bar{M}_3$. In particular we always chose
$\bar{M}_2=5, \bar{M}_3=5, \bar{M}_4=5, \bar{M}_5=5, \bar{M}_6=5$.
Regarding the frequency of each move attempts, we always chose
$g_1=0.001, g_2=1, g_3=1, g_4=1, g_5=1, g_6=1, g_7=0.1$. The
parameter $C$ defining the relative weight of the Z and G sectors
\cite{Fantoni2014b} is adjusted, through short test 
runs, so as to have a Z sector frequency as close as possible to 50\%. 
We accumulate averages over $10^5$ blocks each made of
$10^5$ attempted moves with quantities measured every $10^3$ attempts.
Since the volume move is the most computationally expensive one we
chose its frequency as the lowest. During the simulation we monitor
the acceptance ratios of each move. The various simulations took no
more than $\sim 150$ CPU hours on a 3 GHz processor.

%%%%%%%%%%%%%%%%%%%%%%%%%%%%%%%%%%%%%%%%%%%%%%%%%%%%%%%%%%%%%%%%%%%%%%%%%%%%%%
\subsection{Barotropic phenomenon and gravitational inversion}
%%%%%%%%%%%%%%%%%%%%%%%%%%%%%%%%%%%%%%%%%%%%%%%%%%%%%%%%%%%%%%%%%%%%%%%%%%%%%%
\label{sec:inversion}

The condition for the gravitational inversion observed experimentally
\cite{Sneed1968} is
\bq \nonumber
&&\rho^{(I)}\left(m_\text{He}x_\text{He}^{(I)}+m_{\text{H}_2}x_{\text{H}_2}^{(I)}\right)
>\\ \label{ginv1}
&&\rho^{(II)}\left(m_\text{He}x_\text{He}^{(II)}+m_{\text{H}_2}x_{\text{H}_2}^{(II)}\right),
\eq
where $m_\text{He}/m_{\text{H}_2}=1.98553$. When this condition on the
mass density inversion respect to the number density is satisfied, the
liquid phase will float on top of the vapor phase. The condition of
Eq. (\ref{ginv1}) can also be rewritten as
\bq \label{ginv2}
\rho^{(I)}\left(1+kx_\text{He}^{(I)}\right)>
\rho^{(II)}\left(1+kx_\text{He}^{(II)}\right),
\eq
where $k=m_\text{He}/m_{\text{H}_2}-1=0.98553$. This condition may be
satisfied when the concentration of He in the vapor phase is bigger than
in the liquid phase, at low temperatures, and the number density of the
liquid is close to the one of the vapor, at high pressures. We expect
quantum effects to become important in this regime, before
solidification which is expected to occur for $T<T_s(P)$. The
gravitational inversion of Eq. (\ref{ginv2}) will be satisfied for
$T<T_g(P)$. The experimental $s$-line $T=T_s(P)$ and $g$-line
$T=T_g(P)$ have been determined in Fig. 3 of Sneed {\sl et al.}
\cite{Sneed1968}) in the laboratory.    

%%%%%%%%%%%%%%%%%%%%%%%%%%%%%%%%%%%%%%%%%%%%%%%%%%%%%%%%%%%%%%%%%%%%%%%%%%%%%%
\subsection{Pressure calculation}
%%%%%%%%%%%%%%%%%%%%%%%%%%%%%%%%%%%%%%%%%%%%%%%%%%%%%%%%%%%%%%%%%%%%%%%%%%%%%%
\label{sec:pressure}

We will use the virial estimator for the pressure (see Eq. (6.18) of
Ref. \cite{Ceperley1995}). With long-range corrections
\cite{Allen-Tildesley} which can be quite big in the liquid
phase. More details on the pressure calculation are given in the
supplemental material \cite{Supplemental}.

%%%%%%%%%%%%%%%%%%%%%%%%%%%%%%%%%%%%%%%%%%%%%%%%%%%%%%%%%%%%%%%%%%%%%%%%%%%%%%
\section{Numerical results}
%%%%%%%%%%%%%%%%%%%%%%%%%%%%%%%%%%%%%%%%%%%%%%%%%%%%%%%%%%%%%%%%%%%%%%%%%%%%%%
\label{sec:results}

\begin{figure}[htbp]
\begin{center}
\includegraphics[width=8cm]{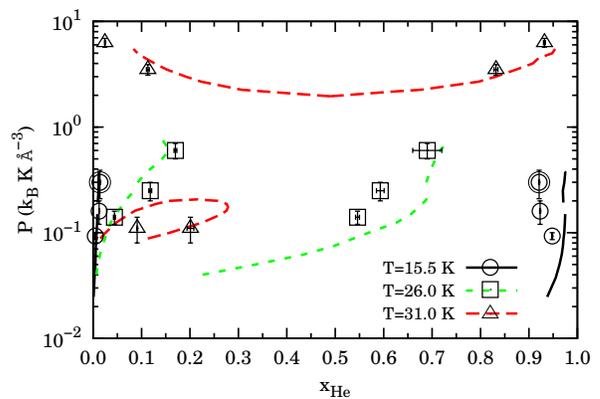}
\end{center}  
\caption{(color online) Comparison between the results of our
  numerical experiments, points from Table \ref{tab:results}, and of
  the laboratory experiments, lines from Table I in the supplemental
  material \cite{Supplemental}, for the pressure-composition of three
  isotherms of the 
  hydrogen-helium mixture phase diagram. A logarithmic scale is
  conveniently used on the ordinates. The double circled points at
  $T=15.5~\text{K}$ denote the case where we observed gravity
  inversion in the numerical experiment.}     
\label{fig:pd}
\end{figure}

Our results are summarized in Table \ref{tab:results} and compared, in
Fig. \ref{fig:pd}, with the experimental data of
Refs. \cite{Streett1964,Sneed1968,Streett1973} (summarized in a Table
in the supplemental material \cite{Supplemental}). 

In all studied cases we chose $N=128$ and $\delta\tau^*=0.002$. We
explored the vapor-liquid coexistence (in this work we will denote with
``vapor-liquid'' coexistence one where $\rho^{(I)}\neq\rho^{(II)}$) at
five temperatures, $T=2, 5, 15.5, 26, 31$ degrees Kelvin, and the
fluid-fluid coexistence (in this work we will denote with
``fluid-fluid'' coexistence one where $\rho^{(I)}=\rho^{(II)}$) at
$T=31~\text{K}$. For the first two lower temperatures studied we could
not find any experimental data for a comparison. In these two cases
when we put a number with trailing dots in the table it means that
after the initial equilibration period the measured property did not
change anymore during the rest of the simulation. 
 
For the temperature $T=15.5~\text{K}$, as it can be readily verified
using the relation of Eq. (\ref{ginv2}), we observe gravitational
inversion on the point at $\rho=0.02~\text{\r{A}}^{-3}$ when the
component with the highest degeneracy temperature is the hydrogen in
the liquid phase with $T_D\approx 2~\text{K}$. Clearly choosing higher
pressures quantum statistics will become more and more important for
the fluid mixture before reaching the solid state. 

For the points at
$T=26~\text{K}$, 
$T=31~\text{K},\rho=0.006~\text{\r{A}}^{-3},\chi=116/12$, and
$T=31~\text{K},\rho=0.03~\text{\r{A}}^{-3},\chi=1$ we 
observed exchanges of identity between the two phases, during the
simulation. 

At a temperature $T=31~\text{K}$ and a pressure of
$P=0.07(2)~k_B\text{K}\text{\r{A}}^{-3}$ we found a vapor-liquid
coexistence, choosing $\chi=116/12$. This point should be subject to
greater size error than all other points simulated, and be thus the
less reliable, since we only have, in the two boxes,
a total of 12 helium atoms. Increasing the pressure to
$P=0.21(2)~k_B\text{K}\text{\r{A}}^{-3}$, in agreement with the
experiment, we did not find coexistence and we observed
$\rho^{(I)}\approx\rho^{(II)}\approx\rho$ and 
$x_\text{He}^{(I)}\approx x_\text{He}^{(II)}\approx
1/(1+\chi)$. Increasing the pressure to
$P=3.5(4)~k_B\text{K}\text{\r{A}}^{-3}$, we did not observe 
exactly $\rho^{(I)}=\rho^{(II)}$, as measured in the fluid-fluid
transition observed in the laboratory \cite{Streett1973}. The same
holds true for the point at the same temperature but higher pressure
$P=6.3(6)~k_B\text{K}\text{\r{A}}^{-3}$.

For all measured points except the one at the lower temperature,
$T=2~\text{K}$ of Table \ref{tab:results}, the superfluid fraction
\cite{Pollock1987} of the two components in either phase was
negligibly small. At $T=2~\text{K}$ of Table \ref{tab:results}, below
the helium lambda-temperature, we observed a negligible superfluid
fraction of both components in the liquid phase and of the hydrogen in
the vapor phase. The helium in the vapor phase was found to have a
superfluid fraction of $0.012(3)$, indicating a tendency to
supefluidity.  

\begin{figure}[htbp]
\begin{center}
\includegraphics[width=8cm]{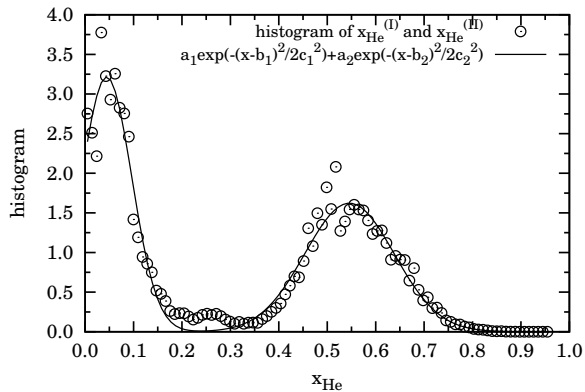}
\end{center}  
\caption{Fit of the histogram for the block averages of both
  $x_{He}^{(I)}$ {\sl and} $x_{He}^{(II)}$ with the sum of two
  gaussians with six parameters. This is case
  $T=26~\text{K},\rho=0.01~\text{\r{A}}^{-3},\chi=90/38$, where we had
  box identity excanges.}  
\label{fig:histogram}
\end{figure}

When we do not observe exchanges of identity between the two phases,
during the simulation, we are able to find accurate average values for
the various measured quantities. 
Otherwise a histogram analysis of the data is necessary with a
non-linear fit using the superpositions of two shifted Gaussians. 
For example in Fig. \ref{fig:histogram} we show the
procedure used to extract the helium concentrations of the two
coexisting phases for the case
$T=26~\text{K},\rho=0.01~\text{\r{A}}^{-3},\chi=90/38$.

The measured property which is less accurate is the pressure due to
the size error and the long-range correction dependent on the
$r_\text{cut}$ choice. This problem could be overcome by using instead
of the $N,V,T$ version of the Gibbs ensemble algorithm its $N,P,T$ one
\cite{Frenkel-Smit}. 

\begin{table*}[htb]
\caption{Numerical isothermal pressure-composition at coexistence. We
  always used $N=128$ and $\delta\tau^*=0.002$.} 
\begin{ruledtabular}
\begin{tabular}{lll||lll|ll}
  $T~(\text{K})$ & 
  $\rho~(\text{\r{A}}^{-3})$ & $\chi$ &
  $P~(k_B\text{K}\text{\r{A}}^{-3})$ &
  $x_\text{He}^{(II)}$ & 
  $x_\text{He}^{(I)}$ & 
  $\rho^{(II)}~(\text{\r{A}}^{-3})$ &
  $\rho^{(I)}~(\text{\r{A}}^{-3})$ \\  
\hline
2.0  & 0.015 & 1     & $-$0.08(7) & 0.214$\ldots$& 0.639$\ldots$& 0.02456(1) & 
0.012605(2)  \\
\hline
5.0  & 0.010 & 1     & 0.014(2) & 0.1787(1) & 1.000$\ldots$   & 0.025910(6)& 
0.005113(1)  \\
\hline
15.5 & 0.010 & 1     & 0.093(7) & 0.00457(9)& 0.948(1)  & 0.02410(1) &
0.006544(5)  \\
15.5 & 0.015 & 1     & 0.16(4)  & 0.0125(3) & 0.923(1)  & 0.02304(2) & 
0.011525(7)  \\
15.5 & 0.020 & 1     & 0.30(9)  & 0.0142(4) & 0.921(1)  & 0.02373(2) &      
0.017619(5)  \\
\hline
26.0 & 0.010 & 90/38 & 0.14(2) & 0.044(2)  & 0.546(4)  & 0.01890(5) & 
0.00669(1)   \\
26.0 & 0.015 & 90/38 & 0.25(5) & 0.118(3) & 0.593(8)  & 0.01888(7) &
0.01105(5)  \\ 
26.0 & 0.020 & 90/38 & 0.6(1) & 0.170(3) & 0.69(3)  & 0.02115(2) &
0.01759(8)  \\ 
\hline
31.0 & 0.006 & 116/12& 0.11(3)  & 0.091(1)  & 0.201(7)  & 0.014(2)   &
0.00564(6)   \\
31.0 & 0.008 & 1     & 0.21(2)  & 0.5025(6) & 0.511(1)  & 0.008016(7) &
0.00795(1)   \\
31.0 & 0.030 & 1     & 3.5(4)   & 0.832(4)  & 0.113(3)  & 0.03198(4) & 
0.02805(1)   \\
31.0 & 0.035 & 1     & 6.3(6)   & 0.932(2)  & 0.0243(9) & 0.03955(5) &
0.03111(1)   \\
\end{tabular}
\end{ruledtabular}
\label{tab:results}
\end{table*}

%%%%%%%%%%%%%%%%%%%%%%%%%%%%%%%%%%%%%%%%%%%%%%%%%%%%%%%%%%%%%%%%%%%%%%%%%%%%%%
\subsection{Finite size effects}
%%%%%%%%%%%%%%%%%%%%%%%%%%%%%%%%%%%%%%%%%%%%%%%%%%%%%%%%%%%%%%%%%%%%%%%%%%%%%%

We studied the finite size effects at
$T=31~\text{K},\rho=0.03~\text{\r{A}}^{-3},\chi=1$. In Table
\ref{tab:size} we show the results for the isothermal
pressure-composition coexistence at $N=64,128,$ and $256$.
As the number of particles increases we observe a decrease in the
ratio of number of exchanges of identity between the two phases and
total number of particles: For $N=64$ the exchanges occurred many times,
for $N=128$ only once, and for $N=256$ never. For the case $N=64$ we
found the peak of the first gaussian for the hystogram of $x_\text{He}$
with a negative value. The simulation with $N=64$ took $1.0\times
10^5$s, the one with $N=128$ took $1.6\times 10^5$s, and the one with
$N=256$ took $4.0\times 10^6$s. From the comparison we see how there is
not much difference between $N=128$ and $N=256$. Apart from the
smaller statistical errors in the latter case, the concentrations
slightly differ in the two cases.

\begin{table*}[htb]
\caption{Numerical isothermal pressure-composition coexistence at
  $T=31~\text{K},\rho=0.03~\text{\r{A}}^{-3},\chi=1$ as a function of
  the number of particles $N$. We always used $\delta\tau^*=0.002$.}  
\begin{ruledtabular}
\begin{tabular}{ll||lll|ll}
  $N$ & $r_\text{cut} (\text{\r{A}})$ &
  $P~(k_B\text{K}\text{\r{A}}^{-3})$ &
  $x_\text{He}^{(II)}$ & 
  $x_\text{He}^{(I)}$ & 
  $\rho^{(II)}~(\text{\r{A}}^{-3})$ &
  $\rho^{(I)}~(\text{\r{A}}^{-3})$ \\  
\hline
64  & 5 & 2.4(8)   & 0.83(3)   & -         & 0.03144(7) & 0.02782(3)\\
128 & 6 & 3.5(4)   & 0.832(4)  & 0.113(3)  & 0.03198(4) & 0.02805(1)\\
256 & 8 & 3.4(2)   & 0.840(3)  & 0.098(3)  & 0.03180(3) & 0.028170(9)\\
\end{tabular}
\end{ruledtabular}
\label{tab:size}
\end{table*}

%%%%%%%%%%%%%%%%%%%%%%%%%%%%%%%%%%%%%%%%%%%%%%%%%%%%%%%%%%%%%%%%%%%%%%%%%%%%%%
\subsection{Importance of the particle exchanges and of the quantum
  effects} 
%%%%%%%%%%%%%%%%%%%%%%%%%%%%%%%%%%%%%%%%%%%%%%%%%%%%%%%%%%%%%%%%%%%%%%%%%%%%%%

Setting to zero the frequency of the swap move attempts our algorithm
reduces to a path integral calculation for distinguishable particles
obeying to the Boltzmann statistics. On the other hand, choosing $K=2$
(with $\bar{M}_q=1$ for all $q$) and
$\lambda^*_{\text{H}_2}=\lambda^*_{\text{He}}\to 0$ it reduces to 
the classical Gibbs Ensemble Monte Carlo (GEMC) algorithm 
of Panagiotopoulos \cite{Panagiotopoulos87}. 

For the state point
$T=15.5~\text{K}$ and $\rho=0.02~\text{\r{A}}^{-3}$ with $N=128$, we
performed two simulations for each of the two cases suggested above to
estimate the importance of particles exchanges which underlies the
Bose-Einstein statistics and of quantum effects, respectively. To
reach the GEMC limit from our QGEMC algorithm we chose, in particular,
$\lambda^*_{\text{H}_2}=\lambda^*_{\text{He}}=10^{-3}$. The 
results are shown in Table \ref{tab:qe}. The 
acceptance ratio for the swap move was around $0.5$ in the full
quantum case and imposed zero in the other two simulations. 

As we can see from the table, for this state point, there is a very
small difference between the path integral simulation with the full
Bose-Einstein statistics and the one with the Boltzmann statistics. In 
particular, only the densities of the vapor phase are different in the
two cases. In both these simulations we observe the gravitational 
inversion. We expect that increasing the pressure and thereby the
density or reducing the temperature the particles exchanges will
become increasingly important. 

On the other hand, there is a large difference between these
two simulations and the classical GEMC one. In particular, the
gravitational inversion is not observed in the classical limit
simulation, even if after a short equilibration time the simulation
converged towards the condition $x_\text{He}^{(I)}=1$, i.e. all helium
atoms, the heaviest species in the mixture, were found in the less
dense phase.

\begin{table*}[htb]
\caption{Numerical isothermal pressure-composition coexistence at
  $T=15.5~\text{K},\rho=0.02~\text{\r{A}}^{-3},\chi=1$ in a simulation
  with the full QGEMC algorithm with the Bose-Einstein statistics
  ($\delta\tau^*=0.002$), with the QGEMC algorithm with Boltzmann
  statistics ($\delta\tau^*=0.002$), and with the GEMC limit (see main
  text) of the QGEMC algorithm. We always used $N=128$.} 
\begin{ruledtabular}
\begin{tabular}{l|lll|ll}
  Statistics &
  $P~(k_B\text{K}\text{\r{A}}^{-3})$ &
  $x_\text{He}^{(II)}$ & 
  $x_\text{He}^{(I)}$ & 
  $\rho^{(II)}~(\text{\r{A}}^{-3})$ &
  $\rho^{(I)}~(\text{\r{A}}^{-3})$ \\  
\hline
QGEMC: Bose-Einstein & 0.30(9)  & 0.0142(4) & 0.921(1)  & 0.02373(2) &
0.017619(5)  \\ 
\hline
QGEMC: Boltzmann & 0.30(9)  & 0.0143(4) & 0.919(1)  & 0.02373(2) &
0.017638(5)  \\ 
\hline
GEMC: classical limit & 0.13(4)  & 0.000$\ldots$ & 1.000$\ldots$  &
0.035953(5) & 0.0138552(7)  \\   
\end{tabular}
\end{ruledtabular}
\label{tab:qe}
\end{table*}

%%%%%%%%%%%%%%%%%%%%%%%%%%%%%%%%%%%%%%%%%%%%%%%%%%%%%%%%%%%%%%%%%%%%%%%%%%%%%%
\section{Conclusions}
%%%%%%%%%%%%%%%%%%%%%%%%%%%%%%%%%%%%%%%%%%%%%%%%%%%%%%%%%%%%%%%%%%%%%%%%%%%%%%
\label{sec:conclusions}

In conclusion, we performed path integral Monte Carlo simulations, using
our newly developed QGEMC method, for the two phase coexistence of the
hydrogen-helium mixture away from freezing. This asymmetric mixture
displays at low temperature, a big concentration 
asymmetry in the two coexisting phases, whereas the densities of the
two phases tend to become equal at high pressure. This is responsible
for a gravitational inversion, where the liquid, the more dense phase,
with an abundance of hydrogen, floats above the vapor, the less dense
phase, with an abundance of helium. In this 
coexistence region of the temperature-pressure diagram, quantum
statistics is expected to play an important role and in our
simulations we are able to observe such gravitational inversion. Our
numerical experiments are also in good quantitative agreement
with the experimental results of C. M. Sneed, W. B. Streett,
R. E. Sonntag, and G. J. Van Wylen in the late 1960's and early
1970's. The difference between our results on the helium concentration
in the two phases and the experimental ones is in all cases less than
15\% in the high helium concntration phase and than 5\% in the low
helium concentration phase, relative to the experiment.

These results for the hydrogen-helium mixture can be of interest for
the study of cold exoplanets with an atmosphere made predominantly by
such a fluid mixture and with the right temperature and 
pressure conditions for there to be coexistence. In such cases it
could be possible to observe the gravitational inversion phenomenon
and consequent changes in the planet moment of inertia, depending on
the atmospherical and climatic conditions. 

At extremely low temperature and pressure we find that the first
component to show superfluidity is the helium in the vapor phase.

Our QGEMC method \cite{Fantoni2014b} is extremely simple to use,
reduces to the Gibbs ensemble method of Panagiotopoulos
\cite{Panagiotopoulos87} in the classical regime, and gives an exact
numerical solution of the statistical physics phase coexistence
problem for boson fluids.

An open problem, currently under exam, is the influence of the
finite-size effects on the determination of the binodal curves close
to the lower strongly asymmetric critical points, as for example in
our case
$T=31~\text{K},\rho=0.006~\text{\r{A}}^{-3},\chi=116/12$. This 
requires additional simulations at an higher and lower number of
particles.     

\appendix
%%%%%%%%%%%%%%%%%%%%%%%%%%%%%%%%%%%%%%%%%%%%%%%%%%%%%%%%%%%%%%%%%%%%%%%%%%%%%%
%\section{...} 
%%%%%%%%%%%%%%%%%%%%%%%%%%%%%%%%%%%%%%%%%%%%%%%%%%%%%%%%%%%%%%%%%%%%%%%%%%%%%%
\label{app:1}

%%%%%%%%%%%%%%%%%%%%%%%%%%%%%%%%%%%%%%%%%%%%%%%%%%%%%%%%%%%%%%%%%%%%%%%%%%%%%% 
%\begin{acknowledgments}

%\end{acknowledgments}
%%%%%%%%%%%%%%%%%%%%%%%%%%%%%%%%%%%%%%%%%%%%%%%%%%%%%%%%%%%%%%%%%%%%%%%%%%%%%%
%\bibliographystyle{apsrev}
\bibliography{qgemc}
%%%%%%%%%%%%%%%%%%%%%%%%%%%%%%%%%%%%%%%%%%%%%%%%%%%%%%%%%%%%%%%%%%%%%%%%%%%%%%
%%%%%%%%%%%%%%%%%%%%%%%%%%%%%%%%%%%%%%%%%%%%%%%%%%%%%%%%%%%%%%%%%%%%%%%%%%%%%%
%%%%%%%%%%%%%%%%%%%%%%%%%%%%%%%%%%%%%%%%%%%%%%%%%%%%%%%%%%%%%%%%%%%%%%%%%%%%%%
\end{document}